\documentclass[printer]{aa}
\usepackage[utf8]{inputenc}
\usepackage{microtype}
\usepackage{paralist}
\usepackage{amsmath}
\usepackage{natbib}
\usepackage{graphicx}
\usepackage{epsfig}
\usepackage{txfonts}
\usepackage{color}
\usepackage{multirow}
\usepackage{morefloats}
\usepackage{amssymb}
\usepackage{epsfig}
\usepackage{xspace}
\bibpunct{(}{)}{;}{a}{}{,} 
\usepackage[plainpages=false]{hyperref}
\newcommand{\xte}{XTE~J1946$+$274\xspace}
\title{\emph{BeppoSAX} observations of \hbox{XTE~J1946$+$274}}
\author{R.\,Doroshenko\inst{1}, A.\,Santangelo\inst{1}, V.\,Doroshenko\inst{1}, S.\,Piraino\inst{2,1}}
\institute{
Institut für Astronomie und Astrophysik, Kepler Center for Astro and Particle Physics, 
Sand 1, 72076 Tübingen, Germany 
\and INAF – IASF di Palermo, via Ugo La Malfa 153, 90146 Palermo, Italy}

\begin{document} 

\bibliographystyle{aa}

\abstract{ We report on the \emph{BeppoSAX} monitoring of a giant outburst of the transient X-ray
pulsar \xte in 1998. The source was detected with a
flux of $\sim4 \cdot 10^{-9}$ erg\,cm$^{-2}$s$^{-1}$
(in $0.1 - 120$~keV range). The broadband spectrum, typical for
accreting pulsars, is well described by a cutoff power law with a cyclotron resonance
scattering feature (CRSF) at $\sim38$~keV. This value is consistent
with earlier reports based on the observations with \emph{Suzaku} at
factor of ten lower luminosity, which implies that the feature is
formed close to the neutron star surface rather than in the accretion
column. Pulsations with $P\sim15.82$\,s were observed up to $\sim70$~keV. The pulse profile strongly depends on energy and is
characterised by a ``soft'' and a ``hard'' peaks shifted by half
period, which suggests a strong phase dependence of the spectrum, and
that two components with roughly orthogonal beam patterns
are responsible for the observed pulse shape. This conclusion is
supported by the fact that the CRSF, despite its relatively high
energy, is only detected in the spectrum of the soft peak of the pulse profile.
Along with the absence of correlation of the line energy with luminosity,
this could be explained in the framework of the recently
proposed ``reflection'' model for CRSF formation. 
However more detailed modelling of both line and continuum formation
are required to confirm this interpretation.}

\authorrunning{R. Doroshenko et al.}
\maketitle

\section{Introduction}

The transient X-ray pulsar \xte was discovered, at a flux level 
F$_{(2 - 12) \rm keV}$ of $\sim90$~mCrab, on September 15, 1998 
by the \emph{All Sky Monitor} (\emph{ASM}) on board the \emph{Rossi X-ray
Timing Explorer} (\emph{RXTE}) \citep{Smith:1998p3835}. From
archival data, it was found that the source had been brightening from $\sim13$
mCrab on September~5 to $\sim 60$~mCrab on September~15. Coherent pulsations
at $\sim 15.83$~s were discovered in follow-up pointed observations by 
\emph{RXTE} \citep{Smith:1998p3835}.
The source was also observed by \emph{BATSE} onboard \emph{CGRO}
\citep{Wilson:1998p3746} which reported a (20 -- 50 keV) flux of $\sim 15$~mCrab.

The pulsar remained active for about three years after that as 
monitored by the \emph{RXTE/ASM}. Following the \emph{RXTE} 
outburst detection, the \emph{BeppoSAX} Target
of Opportunity Observation Program on Hard X--ray Transients 
was activated at F$_{(1 - 10) \rm keV}\sim 44$~mCrab
\citep{Campana:1998p1084}, at the decline of the first, giant, outburst.
The source was also observed
by the \emph{Indian X-ray Astronomy Experiment} (\emph{IXAE}) 
\citep{Paul:2001p3709} that revealed double peaked pulse profiles 
with a pulse fraction of $\sim30\%$ in the 2 -- 6~keV and 
6 -- 18 keV energy band, and confirmed the secular spin-up of
the pulsar. The observed outbursts are most likely associated with
the orbital motion. \cite{Campana:1999p3677} based on \emph{RXTE/ASM} 
data reported evidence for the $\sim80$~d periodicity, 
while \cite{Wilson:2003p2397} confirmed a 169.2\,d orbital period 
based on the X-ray timing measurements.

\cite{Verrecchia:2002p2366} identified the most likely optical
counterpart as a $R \sim 14$ mag \emph{Be} star, that shows a strong
$H_{\alpha}$ emission line. This allowed them to estimate a distance of 8
-- 10 kpc to the source based on the observed extinction. \cite{Wilson:2003p2397},
using evidence for an accretion disc, estimated the distance to be
$d = 9.5 \pm 2.9$ kpc.

Several weaker outbursts followed the first one from 1998 to 2001
\citep{Campana:1999p3677,Paul:2001p3709,Wilson:2003p2397}.
Similar behaviour has been also observed in other
\emph{Be} systems \citep[see for example][] {Caballero:2013p2812}.
The first outburst is, therefore, considered to be a giant one.

 \xte remained in quiescence until June 2010, when the Burst Alert 
 Telescope (\emph{BAT}) on board \emph{Swift} and the Gamma-ray 
 Burst Monitor (\emph{GBM}) on board \emph{Fermi} observed a new 
 outburst \citep{Krimm:2010p2398,2010ATel.2847....1F}. The source 
 has been observed with \emph{INTEGRAL} \citep{2010ATel.2692....1C}, 
 \emph{Swift} and \emph{RXTE} \citep{Muller:2012p1153}, and \emph{Suzaku} 
 \citep{2013ApJ...771...96M,Marcu2015}. Also on this occasion, 
 the source exhibited several outbursts of decreasing intensity, 
 following the first one \citep{Muller:2012p1153}.

Based on the \emph{RXTE} data from the outburst in 1998, the broadband 
X-ray spectrum of the source was found to be similar to that of other 
accreting pulsars, and a cyclotron line near 35~keV, has been reported 
by \cite{Heindl:2001p3667}. However, based on data from the 2010 outburst, the presence of a 
cyclotron line at 35~keV was excluded by \cite{Muller:2012p1153}, 
who suggested possible evidence for a cyclotron feature at $\sim25$~keV. 
On the contrary, based on \emph{Suzaku} data, \cite{2013ApJ...771...96M} 
and \cite{Marcu2015} found marginal evidence for a line near 35-38~keV, 
and no indication of the 25 keV absorption feature. In this paper we present results from the the unpublished analysis of the 1998 observations 
of the source made with \emph{BeppoSAX}, focusing on the timing 
and spectral properties of the source. 

\section{Observations}

The X-ray satellite \emph{BeppoSAX} \citep{Boella:1997p3749} was a programme of
the Italian Space Agency with the participation of the Netherlands Agency for
Aerospace Programs, which operated in the broad 0.1 -- 300\,keV energy band. 
Besides two wide field cameras
\citep[WFCs,][]{Jager:1997p6454}, the scientific payload included four narrow field instruments (NFIs): 
the Medium Energy Concentrator Spectrometers (MECS) operating in the energy range
1 -- 10 keV \citep{Boella:1997p3748}; the Low Energy Concentrator Spectrometer
\citep[LECS, 0.1 -- 10 keV,][]{Parmar:1997p3751}; the High Pressure Gas
Scintillation Proportional Counter \citep[HPGSPC, 4 -- 120 keV,][]{Manzo:1997p3754}; 
and the Phoswich Detector System \citep[PDS, 15 -- 300 keV,][]{Frontera:1997p3756}. 
In this work we have used the LECS (0.1 -- 4 keV), the MECS (2 -- 10 keV), 
the HPGHPC (4.5 -- 34 keV) and the PDS (15 -- 120 keV). 
We used the standard \emph{BeppoSAX} data processing procedure 
described in details in \emph{BeppoSAX} handbook\footnote{http://www.asdc.asi.it/bepposax/}.

\begin{figure}
\center 
\includegraphics[width=0.5\textwidth]{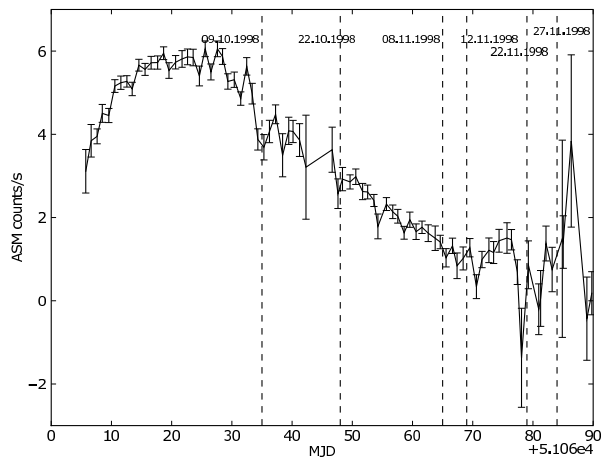}
\caption{Light curve of XTE~J1946$+$274 as observed by RXTE/ASM. Dashed lines are \emph{BeppoSAX} observations.}
\label{asm_lc}
\end{figure}

\xte was observed by \emph{BeppoSAX} in the framework of the AO2 programme
aimed at monitoring the spectral and timing behaviour of hard pulsating 
transients as a function of luminosity. During the decline of the 1998 
giant outburst (Fig. \ref{asm_lc}) of the source, six pointed observations, 
at different luminosities, were performed using the
NFIs (Table \ref{sets}). The analysis of this data has not been published to date.

\begin{table}
\centering
\caption{Observations of the X-ray pulsar \xte by \emph{BeppoSAX}.}
\label{sets}
\begin{tabular}{cccc}
\hline
\hline
Date &  MECS exposure & F$_{unab}^*$ & L$^{**}$\\
& time [ks] & [$10^{-9}$\,erg/cm$^2/$s] &[$10^{37}$\,erg/s]\\
\hline

1998.10.09 & 29 & 4.42 & 4.48\\

1998.10.22 & 23 & 3.15 & 3.4\\

1998.11.08 & 28 & 1.37 & 1.48\\

1998.11.12 & 28 & 1.16 & 1.25\\

1998.11.22 & 12 & 0.63 & 0.68\\

1998.11.27 & 33 & 0.42 & 0.45\\
\hline
\end{tabular}
$*${Unabsorbed fluxes in 0.1 -- 120 keV range.}\\
$**${X-ray luminosity for the distance d$=9.5$ kpc.}\\
\end{table}

\section{Timing analysis}

\begin{figure}
\center 
\includegraphics[width=0.5\textwidth]{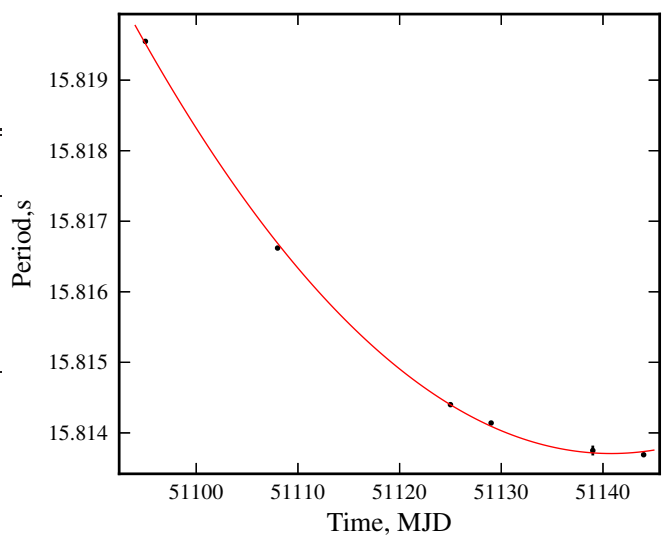}
\caption{XTE~J1946$+$274 pulse period as observed by \emph{BeppoSAX}. Uncertainties
errors are given with $1 \sigma$ confidence level (for the most points errors
are inside the circles).}
\label{period}

\end{figure}

We performed a detailed timing and spectral analysis of all \emph{BeppoSAX} data. For the
timing analysis, the photon arrival times were corrected for motion in the solar
system, and in the binary system assuming ephemeris derived by \cite{Wilson:2003p2397}. To search for pulsation we used epoch folding, and the
obtained period value and uncertainties were refined using the phase-connection
technique. We were not able to find a common timing solution for all
observations assuming a smooth variation of the period, and within individual
observations the pulse arrival times are consistent with constant periods
as listed in Table \ref{perch} (all uncertainties are given at 90\% confidence
level unless stated otherwise). The comparison of the periods measurements
during the outburst reveals a generic spin-up trend as shown in
Fig. \ref{period}, which can be approximated as $P=15.81951(2)$ s, $\dot{P}=-2.99(2)\times10^{-9}$ 
s/s, $\ddot{P}=3.82(4)\times10^{-16}$ s/s$^2$.

\begin{table}
\centering 
\caption{Pulse-period history of XTE~J1946$+$274. For the \emph{BeppoSAX} data, pulsations have been obtained using the phase connection method. Values for \emph{RXTE}, \emph{IXAE} and \emph{Suzaku} are known from \cite{Smith:1998p2184, Paul:2001p3709, Muller:2012p1153, 2013ApJ...771...96M}.}
\label{perch}
\begin{tabular}{ccccc}
\hline
\hline
Date & MJD & Satellite & {P\,[s]} & {$\dot{P}$ [$10^{-9}$\,s/s]} \\
\hline
16.09.1998 & 51072 & \emph{RXTE} & 15.83(2) & ---\\
09.10.1998 & 51095 & \emph{SAX} & 15.81955(4) & 0\\
22.10.1998 & 51108 & \emph{SAX} & 15.81662(4) & 0\\
08.11.1998 & 51125 & \emph{SAX} & 15.81440(4) & 0\\
12.11.1998 & 51129 & \emph{SAX} & 15.81414(4) & 0\\
22.11.1998 & 51139 & \emph{SAX} & 15.8137(2) & 0\\
27.11.1998 & 51144 & \emph{SAX}  & 15.81369(3) & 0\\
24.09.1999 & 51447 & \emph{IXAE} & 15.78801(4) & $-1.54(37)$\\
02.07.2000 & 51727 & \emph{IXAE} & 15.76796(18) & $1.86(49)$\\
20.06.2010 & 55367 & \emph{RXTE} & 15.755(3) & $-3.0(3)$\\
30.06.2010 & 55377 & \emph{RXTE} & 15.767(3) & $-3.0(3)$\\
11.10.2010 & 55480 & \emph{Suzaku} & 15.75(11) & --- \\
\hline
\end{tabular}
\end{table}

\begin{figure}
	\center	
\includegraphics[width=0.5\textwidth]{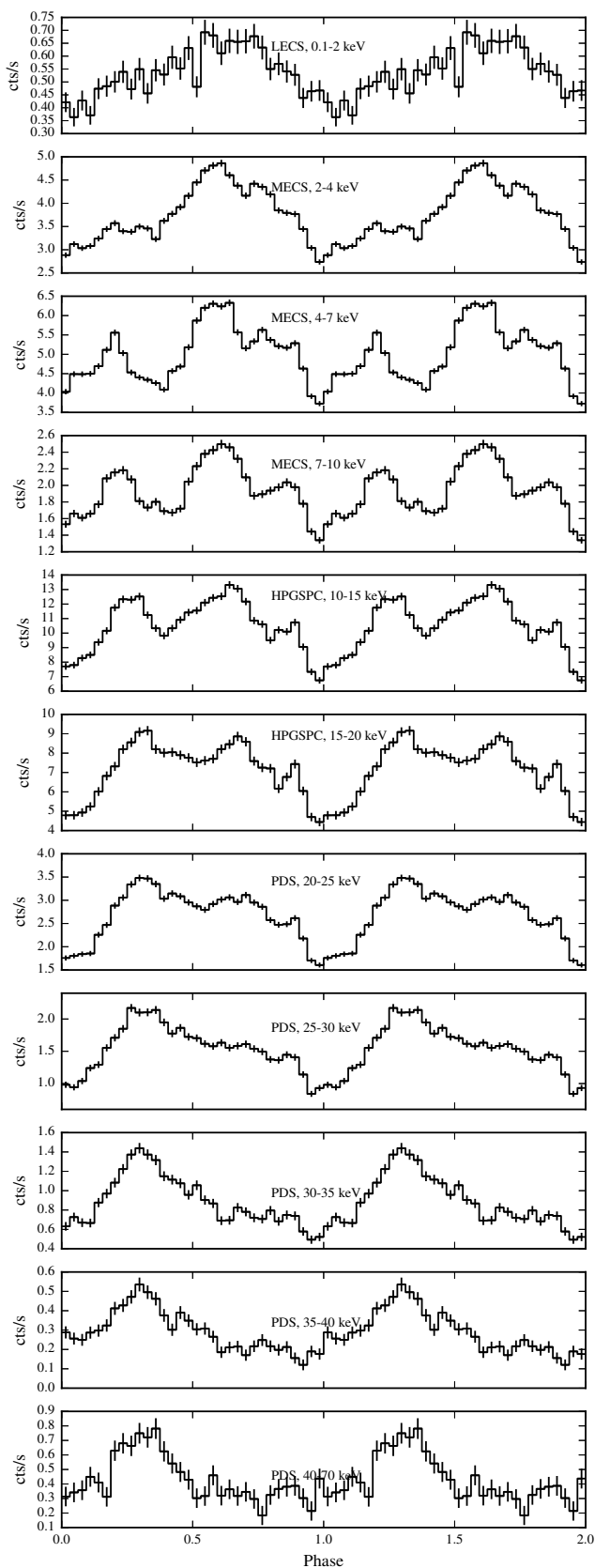}
\caption{The pulse profiles of XTE~J1946$+$274 in 11 energy bands observed with
the \emph{BeppoSAX} satellite by LECS, MECS, HPGSPC and PDS instruments in 9th
of October, 1998.}
\label{pprof}
\end{figure}

The pulse profiles folded with the best-fit period for the brightest
\emph{BeppoSAX} observation are presented in Fig.~\ref{pprof}. The
pulse profile significantly changes with the energy and it is
characterised by two main phase regions: a ``soft'' peak at phases
$\sim(0.5 - 1)$ and a ``hard'' peak 
at $\sim(0 - 0.5)$. The soft peak dominates at low
energies, while the hard peak appears at $\sim1$ keV and steadily
increases with respect to the soft peak till at energies $\ge30$~keV it is the only emission left. Pulsations extend up to $\sim70$~keV. The phase difference between the soft and hard peaks
is about half a phase. 

\begin{figure}
	\center	
\includegraphics[width=0.45\textwidth]{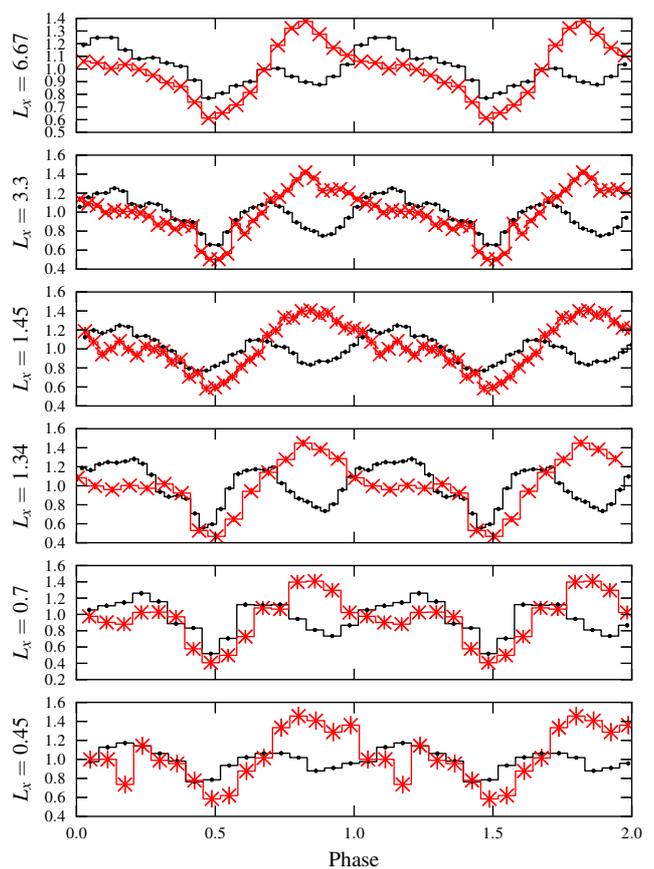}
\caption{The luminosity dependence of the soft (MECS data in 2 -- 10\,keV range, black points),
and hard (PDS data in 20 -- 40\,keV, red crosses) normalised pulse profiles along the outburst.}
\label{pprof_lx}
\end{figure}

The pulse profile shape also changes with
luminosity as can be seen from Fig.~\ref{pprof_lx}. In particular,
both peaks become more pronounced at lower luminosities, while the
shift between the soft and hard peaks remains relatively
constant.

\begin{figure}
\begin{center}  
\includegraphics[width=0.5\textwidth]{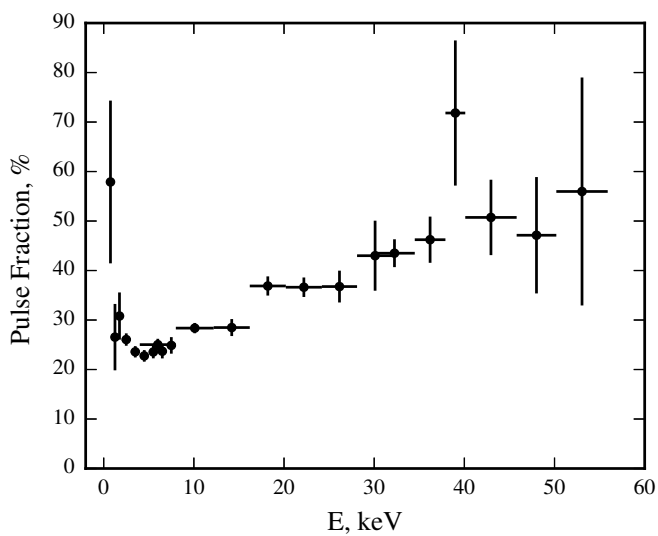}
\caption{\xte pulse fraction for the brightest \emph{BeppoSAX} observation in
10-09-1998. The emission feature is around cyclotron line energy, $E_{cyc} \sim
38$ keV.}
\label{pfrac}
\end{center}
\end{figure}

The fraction of pulsed flux (Fig. \ref{pfrac}), defined as ratio $(F_{max} -
F_{min})/(F_{max} + F_{min})$, where $F_{max}$ and $F_{min}$ are the
maximum and minimum source flux, decreases until $\sim4$\,keV, and then increases 
in 4 -- 60\,keV with a possible feature at $\sim 38$\,keV (see, Section \ref{spe}).

\section{Spectral analysis} 
\label{spe} 
Spectra from the LECS and MECS were extracted from  circles with radii 
8 and 4 arcmin respectively centred on the source, 
while the background was estimated from blank
sky observations. The HPGSPC and PDS were operated in rocking mode, so the
spectra and background were extracted from on- and off-source collimator
positions using the standard \emph{BeppoSAX} pipeline. For fitting the spectra we
used the XSPEC package \citep{Arnaud:1999p2576}.

\begin{figure}
  \begin{center}
        \includegraphics[width=0.5\textwidth]{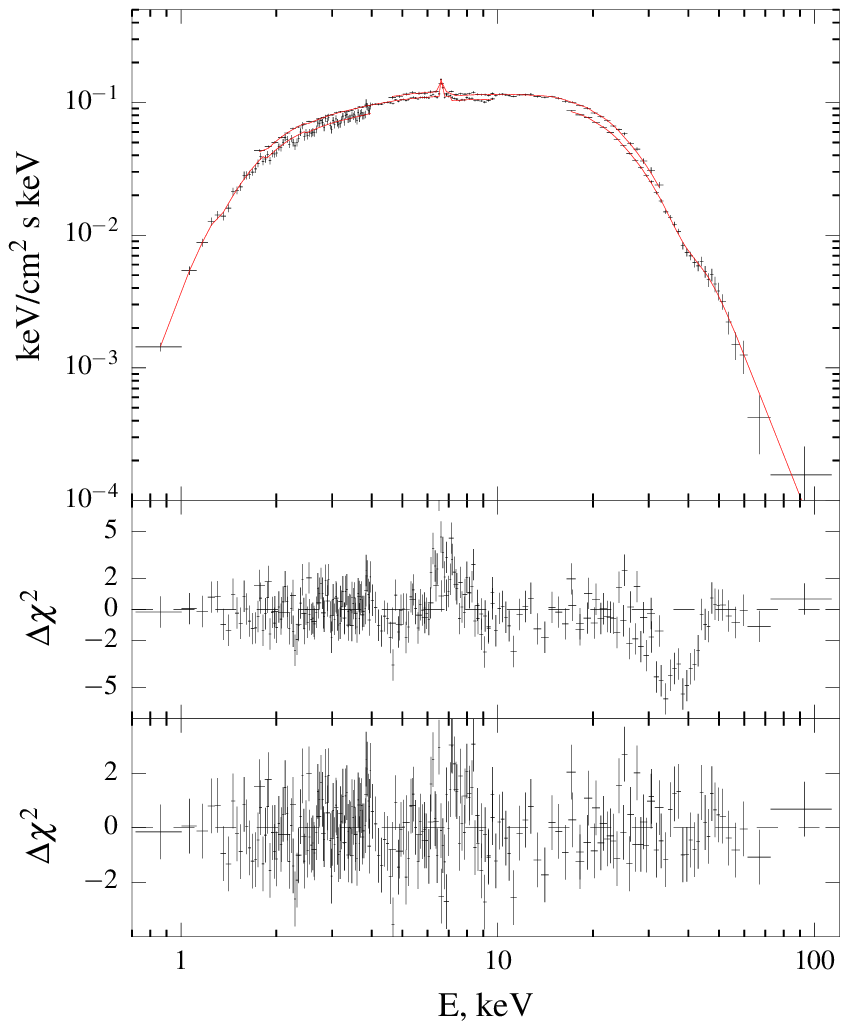}
        \end{center}
        \caption{Unfolded spectrum \xte on 1998-10-09 
        fitted with HIGHECUT+BB continuum model (top panel). The middle and bottom panels show
        best-fit residuals without and with inclusion of iron and cyclotron lines respectively.}
        \label{high_spe}
\end{figure}

The broadband X-ray continuum of the pulse-phase averaged spectrum 
can be described with an absorbed power law cutoff
at high energies. In addition, an iron line at
$\sim6.4-6.6$\, keV, and an absorption line at $\sim 38$\, keV for the Cyclotron
Resonant Scattering Feature (CRSF) are required (Fig. \ref{high_spe}) by the data.

\begin{table*}
\center
\caption{Variation parameters of \xte with flux, during decay of the 1998 yr outburst observed by \emph{BeppoSAX} with HIGHECUT+BB continuum model.}
{\begin{tabular}
{lllllll}
\hline
Parameter &1998.10.09&1998.10.22&1998.11.08&1998.11.12&1998.11.22&1998.11.27\\
\hline
\hline
N$_{\rm H}$ [$10^{22}$atoms cm$^{-2}$] & 
        $1.32_{-0.06}^{+0.07}$&
        $1.12_{-0.1}^{+0.12}$&
        $0.92_{-0.08}^{+0.16}$&
        $1.21_{-0.17}^{+0.16}$&
        $1.27_{-0.54}^{+0.47}$&
        $1.05_{-0.17}^{+0.26}$\\
$\Gamma$&
        $0.88_{-0.03}^{+0.03}$&
        $0.82_{-0.05}^{+0.05}$&
        $0.8_{-0.04}^{+0.08}$&
        $0.92_{-0.11}^{+0.06}$&
        $1.17_{-0.41}^{+0.21}$&
        $0.98_{-0.11}^{+0.14}$\\
E$_{\rm cut}$ [keV]& 
        $18.04_{-0.35}^{+0.53}$&
        $18.45_{-0.5}^{+0.84}$&
        $19.37_{-0.95}^{+2.51}$&
        $19.16_{-1.97}^{+1.16}$&
        $26.84_{-10.16}^{+6.57}$&
        $18.54_{-5.98}^{+7.42}$\\
E$_{\rm fold}$ [keV]& 
        $9.36_{-0.54}^{+0.36}$&
        $8.9_{-0.53}^{+0.38}$&
        $8.95_{-0.68}^{+0.43}$&
        $9.8_{-0.9}^{+0.8}$&
        $9.3_{-3.4}^{+4.6}$&
        $10.35_{-2.27}^{+1.06}$\\
E$_{\rm Fe}^a$ [keV]&
        $6.62_{-0.05}^{+0.05}$&
        $6.66_{-0.09}^{+0.09}$&
        $6.6_{-0.1}^{+0.1}$&
        $6.6_{-0.1}^{+0.1}$&
        $6.62_{-0.05}^{+0.05}$&
        $6.4_{-0.1}^{+0.1}$\\
A$_{\rm Fe}$ [$10^{-3}$ph cm$^{-2}$s$^{-1}$]& 
        $1.1_{-0.2}^{+0.1}$&
        $0.7_{-0.2}^{+0.2}$&
        $0.3_{-0.1}^{+0.1}$&
        $0.2_{-0.1}^{+0.1}$&
        $0.05_{-0.01}^{+0.01}$&
        $0.15_{-0.05}^{+0.05}$\\
E$_{\rm cyc}$ [keV]&
        $38.34_{-1.3}^{+1.45}$&
        $40.79_{-1.92}^{+2.2}$&
        $36.0^e$&
        $38.33_{-1.95}^{+2.76}$&
        $38.0^e$&
        $35.17_{-1.78}^{+2.51}$\\
$\sigma_{\rm cyc}$ [keV]&
        $4.55_{-1.21}^{+1.35}$&
        $4.0^e$&
        $4.0^e$&
        $0.9_{-0.8}^{+2.8}$&
        $4.0^e$&
        $0.4_{-0.3}^{+2.9}$\\
$\tau_{\rm cyc}^b$&
        $0.3_{-0.1}^{+0.1}$&
        $0.3_{-0.2}^{+0.1}$&
        $0.12_{-0.01}^{+0.01}$&
        $1.1_{-0.9}^{+0.9}$&
        $0.59_{-0.01}^{+0.01}$&
        $0.3_{-0.1}^{+0.7}$\\
kT$_{\rm bb}$ [keV]& 
        $1.87_{-0.11}^{+0.12}$&
        $1.96_{-0.12}^{+0.18}$&
        $2.12_{-0.12}^{+0.16}$&
        $2.37_{-0.28}^{+0.23}$&
        $2.4_{-0.43}^{+0.27}$&
        $1.98_{-0.28}^{+0.21}$\\
R$_{\rm bb}^c$&
        $1.4_{-0.4}^{+0.8}$&
        $1.2_{-0.6}^{+0.7}$&
        $0.9_{-0.4}^{+0.4}$&
        $0.7_{-0.2}^{+0.4}$&
        $0.7_{-0.4}^{+0.4}$&
        $0.7_{-0.3}^{+0.4}$\\
F$_{\rm ab}^d$&
        $4.13_{-0.03}^{+0.01}$&
        $2.98_{-0.03}^{+0.02}$&
        $1.31_{-0.05}^{+0.01}$&
        $1.09_{-0.03}^{+0.02}$&
        $0.58_{-0.06}^{+0.04}$&
        $0.41_{-0.13}^{+0.01}$\\
F$_{\rm unab}^d$&
        $4.42_{-0.07}^{+0.01}$&
        $3.15_{-0.07}^{+0.01}$&
        $1.37_{-0.07}^{+0.01}$&
        $1.16_{-0.05}^{+0.01}$&
        $0.63_{-0.08}^{+0.03}$&
        $0.42_{-0.08}^{+0.04}$\\
\hline
$\chi^2_{\rm res}$ / dof &
        1.065 / 599 &
        1.104 / 151 &
        0.971 / 283 &
        0.91 / 173 &
        0.888 / 63 &
        0.894 / 212 \\
\hline
\end{tabular}}

$^a${$\sigma_{\rm Fe} = 0.01$.}\\
$^b${$\tau_{\rm cyc}$ is the optical depth.}\\
$^c${R$_{\rm bb}$ is the radius of the black body in km for the distance to the source D$ = 9.5$ kpc.}\\
$^d${The values of absorbed and unabsorbed fluxes in 0.1 -- 120 keV energy range in units $10^{-9}$erg / cm$^2/$s.}\\
$^e${This parameter was fixed.}
\label{xte_high_alldata}
\end{table*}

Following the literature, we considered several models for the continuum. A
power law with a high-energy cutoff and a smoothing gaussian at
cutoff energy \citep{White1983,Coburn2002}; a power law with a Fermi-Dirac cutoff \citep{Tanaka1986}; the negative
and positive power law exponential model \citep{Mihara:1995p3766,
Makishima:1999p3771} and the Comptonisation model by
\cite{Titarchuk:1994p6894} (HIGHECUT, FDCUT, NPEX and
CompTT respectively in XSPEC).
For all considered continuum models the addition of an iron line at
$\sim6.4-6.6$\, keV and of an absorption line with Gaussian optical 
depth profile \citep{Hemphill13} at $\sim38$\,keV was necessary. 

To account for the residuals around 10\,keV, similar to those
reported by \cite{Muller:2012p1153}, either a partial covering 
absorber (PC) or another absorption line at
$\sim10$\,keV (with the width of $\sim2$\,keV) was required. 
The quality of the fit was comparable for all models tested. We have also found that
the PC might be substituted with additional
blackbody-like component (BB) with temperature of $\sim2$\,keV. 

The parameters of the CRSF were best constrained when the continuum was
fitted with the HIGHECUT model. We use this model below
(see Fig.~\ref{high_spe}, Tab.~\ref{xte_high_alldata})
for all observations.

The best-fit parameters for the three different models: 
with the PC and HIGHECUT (i.e. the same as used by
\citealt{2013ApJ...771...96M}); with HIGHECUT and BB; 
and with FDCUT and additional absorption feature 
(GAUS at E$_{G} = 10.2(5)$\,keV with $\sigma_{G} = 2.1(5)$\,keV 
i.e. the same model as reported by
\citealt{Muller:2012p1153}), are presented in Table~\ref{mo1}. 

We note that the CRSF parameters are consistent within uncertainties for
all models. The cyclotron line parameters are significantly 
constrained in the first of the \emph{BeppoSAX} observations. 
The feature is required also for the second observation, however, the line width
becomes unconstrained. Following \cite{2013ApJ...771...96M}, who reported that the line
remains narrow at fluxes comparable with the dimmest \emph{BeppoSAX}
observations, for the second observation we fixed the line width
to the same value of the first observation, which
resulted in slightly higher value of the CRSF energy
$E_{cyc}=41(2)$\,keV.

We have also performed pulse-phase resolved analysis for the
brightest observation. To describe the phase resolved spectra, 
we used a HIGHECUT model, modified by partial covering or blackbody. 
We have found that at lower counting statistics the parameters of the 
CRSF are better constrained when the additional blackbody component is used. 

\begin{figure}
        \centering
                \includegraphics[width=0.5\textwidth]{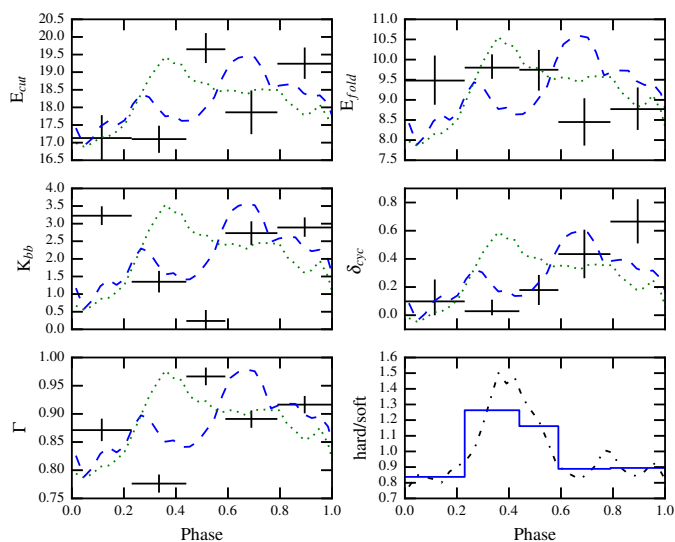}
        \caption{Phase dependence of the spectral parameters for the brightest 
        observation with HIGHECUT+BB model. Blue dashed and green dotted lines 
        show for reference the MECS (2 -- 10 keV) and PDS (15 -- 120 keV) pulse 
        profiles, scaled to match the respective parameter range. The lower right panel shows the hardness ratio for the same energy bands.}
        \label{phres}
\end{figure}

Taking into account the observed pulse profile morphology and significant statistics, we divided the data in five phase bins.  
In the case with the BB we fixed the absorption, energy and width of the emission iron and cyclotron lines
and the temperature of the blackbody component at the average values. The
phase dependence of the best fit parameters is shown in Fig.~\ref{phres}. It
is interesting to note that the depth of the cyclotron feature becomes
consistent with zero at the first pulse phases, meaning that the CRSF is statistically
not significant. In fact, it is only significant in the soft peak of the
pulse profile, even if the counting statistics is lower at this phase.
We also note that the for the third phase bin the spectrum is consistent 
with a pure power law, so the additional soft component
or partial covering are not required at this phase.

\begin{table}
\center
\caption{Parameters of the spectrum in 1998-10-09 with different description of the continuum:
HIGHECUT + BB, HIGHECUT + PC and FDCUT + GAUS.
The values of absorbed and unabsorbed fluxes in 0.1 -- 120 keV energy range are 
F$_{ab} = 4.04 \times 10^{-9}$ erg / cm$^2/$s, F$_{unab} = 4.27 \times 10^{-9}$ erg / cm$^2/$s.
}
\label{mo1}
\begin{tabular}{llll}

\hline
Parameter & \texttt{HC+BB} & \texttt{HC+PC} & \texttt{FDC+G}\\
\hline
\hline
N$_{\rm H}$ [$10^{22}$atoms cm$^{-2}$]& 
        $1.32_{-0.06}^{+0.07}$ & $1.63_{-0.11}^{+0.05}$ & $1.32_{-0.06}^{+0.06}$\\
$\Gamma$ & 
        $0.88_{-0.03}^{+0.03}$ & $1.14_{-0.04}^{+0.02}$ & $0.7_{-0.05}^{+0.04}$\\
E$_{\rm cut}$ [keV]& 
        $18.04_{-0.34}^{+0.53}$ & $18.57_{-0.17}^{+0.4}$ & $16.3_{-1.98}^{+1.32}$\\
E$_{\rm fold}$ [keV]& 
        $9.4_{-0.5}^{+0.4}$ & $9.98_{-0.42}^{+0.41}$ &    $8.1_{-0.3}^{+0.3}$\\
E$_{\rm Fe}^a$ [keV]& 
        $6.62_{-0.05}^{+0.05}$ & $6.66_{-0.06}^{+0.05}$ & $6.61_{-0.05}^{+0.05}$\\
A$_{\rm Fe}$ [$10^{-3}$ph cm$^{-2}$s$^{-1}$]& 
        $1.1_{-0.2}^{+0.1}$ & $1.01_{-0.16}^{+0.22}$ &  $1.2_{-0.2}^{+0.2}$\\
E$_{\rm cyc}$ [keV]& 
        $38.34_{-1.3}^{+1.45}$ & $38.47_{-1.43}^{+1.54}$ & $37.88_{-1.28}^{+1.36}$\\
$\sigma_{\rm cyc}$ [keV]& 
        $4.55_{-1.21}^{+1.35}$ & $4.6_{-1.3}^{+1.5}$ & $7.22_{-1.38}^{+1.08}$\\
$\tau_{\rm cyc}$ &
        $0.30_{-0.08}^{+0.08}$ & $0.27_{-0.08}^{+0.08}$ & $0.5_{-0.1}^{+0.1}$\\
kT$_{\rm bb}$ [keV]& 
        $1.87_{-0.11}^{+0.12}$ & --- & ---\\
K$_{\rm bb}^b$ & 
        $2.2_{-0.5}^{+0.6}$ & --- & ---\\
N$_{\rm H_{pc}}$ [$10^{22}$atoms cm$^{-2}$]& 
        --- &  $12.53_{-3.57}^{+1.72}$ & ---\\
C$_{\rm pc}$ & 
        --- & $0.27_{-0.03}^{+0.03}$ & ---\\
E$_{\rm G}$ [keV]& 
        --- & --- & $10.2_{-0.5}^{+0.3}$ \\
$\sigma_{\rm G}$ [keV]& 
        --- & --- & $2.1_{-0.5}^{+0.5}$ \\
$\delta_{\rm G}$ &
         & & $0.09_{-0.01}^{+0.02}$\\
\hline
$\chi^2_{\rm red}$ / dof & 
        1.065/599 & 1.122/599 & 1.034/599\\
\hline
\end{tabular}
$^a${Width of the iron line fixed at $\sigma = 0.01$ keV.}\\
$^b${$K_{bb} = R^2/D_{10}^2$, where $R^2$ is the black body radius in km,
D$_{10}^2$ is the distance to the source in units of 10 kpc. For the \xte D$ = 9.5$ kpc.}\\
\end{table}

\section{Discussion} 

In this paper we have analysed observations of \xte performed with  
\emph{BeppoSAX}  during the October -- November 1998 outburst of the source. We found that the broadband spectrum of the source has a complex shape,
which is not described adequately by the typically used phenomenological models,
particularly in the energy region around $\sim10$\,keV. This is indeed in agreement with 
what reported by \cite{Muller:2012p1153} who, analysed observations of the 
2010/2011 outburst of the source taken with \emph{INTEGRAL}, \emph{RXTE} and  
\emph{Swift}.  \cite{Muller:2012p1153} accounted for the residuals around 
$\sim10$\,keV by including additional absorption line. However, based on 
the broadband spectra of accreting pulsars observed by \emph{BeppoSAX} 
\citep[see for example][]{Robba2001,Doroshenko2015}, we argue that residuals 
around 10\,keV are due to an incorrect  modelling of the entire continuum 
rather than to some local physical feature. 

In fact, following \cite{2013ApJ...771...96M} who analysed \emph{Suzaku} 
observations of the 2010 outburst of the source, we successfully modelled 
the broadband \emph{BeppoSAX} spectrum of \xte using an additional partial 
covering absorber in combination with the widely used phenomenological 
model HIGHECUT. This implies that part of the emission is  absorbed within 
the binary system. To obtain a statistically acceptable fit, a more elaborate 
modelling of the soft X-ray absorption has been used by \cite{Marcu2015}, 
who, analysing a \emph{Suzaku} observation of the source at lower luminosity, 
considered absorption by different components of the ISM.

In our analysis we have found evidence of a cyclotron line at $\sim38$\,keV. 
The observed line energy implies a value of the magnetic 
field of $3.3(1+z)\times 10^{12}$ G assuming E$_{\rm cyc}~\approx~11.56~\rm B_{12}~/~(1+z)$, 
where $z$ is the gravitational redshift of the scattering region. 
As seen in Table \ref{mo1}, our result is independent of 
the continuum model used. 

Our finding is consistent within 3-sigma with the value reported by \cite{Heindl:2001p3667} 
on the basis of \emph{RXTE} data of the same outburst, but at the higher luminosity. 
Our result is also in agreement with the value of the CRSF energy reported 
by \cite{2013ApJ...771...96M} with the \emph{Suzaku}. 
We note, that the considered \emph{BeppoSAX} observation had been performed 
at significantly higher luminosity, than the \emph{Suzaku} observations. 
So there appears to be no strong correlation of the CRSF centroid energy with luminosity.
On the other hand, the centroid of the not very significant ($\sim2.8\sigma$) 
line at $\sim35$\,keV reported by \cite{Marcu2015} is slightly lower.  
In our analysis we do not find evidence of any line at $\sim 25$\,keV as 
suggested by \cite{Muller:2012p1153}, even if we use the same continuum 
used by \cite{Muller:2012p1153}. Since the source had comparable luminosity 
in the two observations, the reason for such a dramatic change is unclear. 
Conversely, \cite{Muller:2012p1153} report that they finds two 
alternative fit solutions with lines at $\sim30$ and $\sim40$\,keV, and 
while the solution with a line at 25\,keV is most significant of the three, its
significance remains low at $\le2\sigma$. 

A suggestion of the CRSF being located at 38~keV follows 
from the pulsed fraction energy dependance which exhibits
a feature around $\sim40$\,keV (see Fig.~\ref{pfrac}).
Although the feature is marginally significant, we observe that a similar behaviour has also
been observed in other sources and can be associated with change of the pulsar beam
pattern around the resonance energy \citep{Ferrigno:2009p1181,Schoenherr2014}.

We have also carried out the pulse-phase resolved analysis for the
brightest \emph{BeppoSAX} observation. Surprisingly, the CRSF is only
detected at certain pulse phases, namely in the second peak of the two
peaked pulse profile. The statistical quality of the phase-resolved spectra is
comparable in all phase bins, so the feature does indeed change 
intensity and energy with pulse phase. This could be understood considering 
again the pulse profiles of the source. 

As already discussed, the pulse profiles observed by \emph{BeppoSAX} strongly
evolve with the energy (see Fig.~\ref{pprof}). The two peaks are separated 
by about half a phase, and, therefore, likely
represent two emission components with roughly orthogonal beam
patterns. \cite{Paul:2001p3709} suggested that the two peaks might be
related to the emission from the two poles of the neutron star not
diametrically positioned to each other. On the other hand, the pulse
profile evolution, and in particular the $\sim 180^{\circ}$ phase
shift between the soft and hard peaks can be qualitatively
explained in terms of simultaneous presence of a ``hard pencil''
(aligned with cylindrical symmetry about the dipole axis) and a
``soft fan'' (cylindrically symmetric around the axis perpendicular
to the dipole axis) beam components \citep{Woo:1996p1109}. Provided
that $\Theta + \Phi < 90^{\circ}$ ($\Theta$ is the inclination of the
spin axis, $\Phi$ is the angle between the spin and dipole axis) a
single-peak pulse profile is produced by either pencil and fan beams.

The absence of the CRSF at certain phases might be explained,
therefore, if only one of the components exhibits a CRSF in its
spectrum. A similar scenario is realised in the model
proposed by \cite{poutanen} where reflection off the surface
of the neutron star is responsible for CRSF formation. Here we would
like to note that the reflected component is likely to be softer than
incident emission, and the CRSF in \xte is only detected in
soft peak. The fact that the CRSF parameters remain essentially
unchanged despite an order of magnitude change in luminosity, which
was already noted  previously by \cite{Marcu2015}, also suggests that the
CRSF shall be formed close to the surface of the neutron star, rather
than in a tall accretion column. This can be either due to reflection, or
simply because the source consistently accretes in the subcritical regime as suggested by \cite{Marcu2015}.
A detailed modelling of the pulse profile, X-ray continuum and CRSF formation is required to
move from the suggested qualitative interpretation to unambiguous and
quantitative description of the observed phase and luminosity of the
spectrum. This modelling must take into account several effects, from local beam pattern formation to propagation to the observer, which includes gravitational light bending. This is, however, an extremely complex task \citep{Kraus01,Kraus03,Sasaki12}
which is out of the scope of the present paper.

\begin{acknowledgements}
Authors thank the anonymous referee for the helpful comments.
This work is partially supported by the \textsl{Bundesministerium f\"{u}r Wirtschaft und Technologie} through the \textsl{Deutsches Zentrum f\"{u}r Luft- und Raumfahrt e.V. (DLR)} under the grant numbers 50 OG 1301, 50 OR 1310.
\end{acknowledgements}

\bibliography{bib}

\begin{thebibliography}{39}
\expandafter\ifx\csname natexlab\endcsname\relax\def\natexlab#1{#1}\fi

\bibitem[{Arnaud {et~al.}(1999)Arnaud, Dorman, \& Gordon}]{Arnaud:1999p2576}
Arnaud, K., Dorman, B., \& Gordon, C. 1999, Astrophysics Source Code Library,
  10005

\bibitem[{Boella {et~al.}(1997{\natexlab{a}})Boella, Butler, Perola, Piro,
  Scarsi, \& Bleeker}]{Boella:1997p3749}
Boella, G., Butler, R.~C., Perola, G.~C., {et~al.} 1997{\natexlab{a}}, A {\&} A
  Supplement series, 122, 299

\bibitem[{Boella {et~al.}(1997{\natexlab{b}})Boella, Chiappetti, Conti,
  Cusumano, del Sordo, Rosa, Maccarone, Mineo, Molendi, Re, Sacco, \&
  Tripiciano}]{Boella:1997p3748}
Boella, G., Chiappetti, L., Conti, G., {et~al.} 1997{\natexlab{b}}, A {\&} A
  Supplement series, 122, 327

\bibitem[{{Caballero} {et~al.}(2010){Caballero}, {Pottschmidt}, {Bozzo},
  {Ferrigno}, {Neronov}, {Santangelo}, {Klochkov}, {Staubert}, {Kretschmar},
  {Wilms}, {Kreykenbohm}, {Fuerst}, {Schoenherr}, {Rothschild}, \&
  {Suchy}}]{2010ATel.2692....1C}
{Caballero}, I., {Pottschmidt}, K., {Bozzo}, E., {et~al.} 2010, The
  Astronomer's Telegram, 2692

\bibitem[{Caballero {et~al.}(2013)Caballero, Pottschmidt, Marcu, Barragan,
  Ferrigno, Klochkov, Heras, Suchy, Wilms, Kretschmar, Santangelo, Kreykenbohm,
  F{\"u}rst, Rothschild, Staubert, Finger, Camero-Arranz, Makishima, Enoto,
  Iwakiri, \& Terada}]{Caballero:2013p2812}
Caballero, I., Pottschmidt, K., Marcu, D.~M., {et~al.} 2013, The Astrophysical
  Journal Letters, 764, L23

\bibitem[{Campana {et~al.}(1999)Campana, Israel, \& Stella}]{Campana:1999p3677}
Campana, S., Israel, G., \& Stella, L. 1999, Astronomy and Astrophysics, 352,
  L91

\bibitem[{Campana {et~al.}(1998)Campana, Israel, Stella, \&
  Santangelo}]{Campana:1998p1084}
Campana, S., Israel, G.~L., Stella, L., \& Santangelo, A. 1998, IAU Circ.,
  7039, 2

\bibitem[{{Coburn} {et~al.}(2002){Coburn}, {Heindl}, {Rothschild}, {Gruber},
  {Kreykenbohm}, {Wilms}, {Kretschmar}, \& {Staubert}}]{Coburn2002}
{Coburn}, W., {Heindl}, W.~A., {Rothschild}, R.~E., {et~al.} 2002, \apj, 580,
  394

\bibitem[{{Doroshenko} {et~al.}(2015){Doroshenko}, {Santangelo}, {Doroshenko},
  {Suleimanov}, \& {Piraino}}]{Doroshenko2015}
{Doroshenko}, R., {Santangelo}, A., {Doroshenko}, V., {Suleimanov}, V., \&
  {Piraino}, S. 2015, \mnras, 452, 2490

\bibitem[{Ferrigno {et~al.}(2009)Ferrigno, Becker, Segreto, Mineo, \&
  Santangelo}]{Ferrigno:2009p1181}
Ferrigno, C., Becker, P.~A., Segreto, A., Mineo, T., \& Santangelo, A. 2009,
  Astronomy and Astrophysics, 498, 825

\bibitem[{{Finger}(2010)}]{2010ATel.2847....1F}
{Finger}, M.~H. 2010, The Astronomer's Telegram, 2847

\bibitem[{Frontera {et~al.}(1997)Frontera, Costa, dal Fiume, Feroci, Nicastro,
  Orlandini, Palazzi, \& Zavattini}]{Frontera:1997p3756}
Frontera, F., Costa, E., dal Fiume, D., {et~al.} 1997, A {\&} A Supplement
  series, 122, 357

\bibitem[{Heindl {et~al.}(2001)Heindl, Coburn, Gruber, Rothschild, Kreykenbohm,
  Wilms, \& Staubert}]{Heindl:2001p3667}
Heindl, W.~A., Coburn, W., Gruber, D.~E., {et~al.} 2001, The Astrophysical
  Journal, 563, L35, (c) 2001: The American Astronomical Society

\bibitem[{{Hemphill} {et~al.}(2013){Hemphill}, {Rothschild}, {Caballero},
  {Pottschmidt}, {K{\"u}hnel}, {F{\"u}rst}, \& {Wilms}}]{Hemphill13}
{Hemphill}, P.~B., {Rothschild}, R.~E., {Caballero}, I., {et~al.} 2013, \apj,
  777, 61

\bibitem[{Jager {et~al.}(1997)Jager, Mels, Brinkman, Galama, Goulooze, Heise,
  Lowes, Muller, Naber, Rook, Schuurhof, Schuurmans, \&
  Wiersma}]{Jager:1997p6454}
Jager, R., Mels, W.~A., Brinkman, A.~C., {et~al.} 1997, A {\&} A Supplement
  series, 125, 557

\bibitem[{{Kraus}(2001)}]{Kraus01}
{Kraus}, U. 2001, \apj, 563, 289

\bibitem[{{Kraus} {et~al.}(2003){Kraus}, {Zahn}, {Weth}, \& {Ruder}}]{Kraus03}
{Kraus}, U., {Zahn}, C., {Weth}, C., \& {Ruder}, H. 2003, \apj, 590, 424

\bibitem[{Krimm {et~al.}(2010)Krimm, Barthelmy, Baumgartner, Cummings,
  Fenimore, Gehrels, Markwardt, Palmer, Sakamoto, Skinner, Stamatikos, Tueller,
  \& Ukwatta}]{Krimm:2010p2398}
Krimm, H.~A., Barthelmy, S.~D., Baumgartner, W., {et~al.} 2010, The
  Astronomer's Telegram, 2663, 1

\bibitem[{{Maitra} \& {Paul}(2013)}]{2013ApJ...771...96M}
{Maitra}, C. \& {Paul}, B. 2013, \apj, 771, 96

\bibitem[{Makishima {et~al.}(1999)Makishima, Mihara, Nagase, \&
  Tanaka}]{Makishima:1999p3771}
Makishima, K., Mihara, T., Nagase, F., \& Tanaka, Y. 1999, The Astrophysical
  Journal, 525, 978, (c) 1999: The American Astronomical Society

\bibitem[{Manzo {et~al.}(1997)Manzo, Giarrusso, Santangelo, Ciralli, Fazio,
  Piraino, \& Segreto}]{Manzo:1997p3754}
Manzo, G., Giarrusso, S., Santangelo, A., {et~al.} 1997, A {\&} A Supplement
  series, 122, 341

\bibitem[{{Marcu-Cheatham} {et~al.}(2015){Marcu-Cheatham}, {Pottschmidt},
  {K{\"u}hnel}, {M{\"u}ller}, {Falkner}, {Caballero}, {Finger}, {Jenke},
  {Wilson-Hodge}, {F{\"u}rst}, {Grinberg}, {Hemphill}, {Kreykenbohm},
  {Klochkov}, {Rothschild}, {Terada}, {Enoto}, {Iwakiri}, {Wolff}, {Becker},
  {Wood}, \& {Wilms}}]{Marcu2015}
{Marcu-Cheatham}, D.~M., {Pottschmidt}, K., {K{\"u}hnel}, M., {et~al.} 2015,
  \apj, 815, 44

\bibitem[{Mihara(1995)}]{Mihara:1995p3766}
Mihara, T. 1995, Ph.D. thesis, 215

\bibitem[{M{\"u}ller {et~al.}(2012)M{\"u}ller, K{\"u}hnel, Caballero,
  Pottschmidt, F{\"u}rst, Kreykenbohm, Sagredo, Obst, Wilms, Ferrigno,
  Rothschild, \& Staubert}]{Muller:2012p1153}
M{\"u}ller, S., K{\"u}hnel, M., Caballero, I., {et~al.} 2012, Astronomy {\&}
  Astrophysics, 546, 125

\bibitem[{Parmar {et~al.}(1997)Parmar, Martin, Bavdaz, Favata, Kuulkers,
  Vacanti, Lammers, Peacock, \& Taylor}]{Parmar:1997p3751}
Parmar, A.~N., Martin, D. D.~E., Bavdaz, M., {et~al.} 1997, A {\&} A Supplement
  series, 122, 309

\bibitem[{Paul {et~al.}(2001)Paul, Agrawal, Mukerjee, Rao, Seetha, \&
  Kasturirangan}]{Paul:2001p3709}
Paul, B., Agrawal, P.~C., Mukerjee, K., {et~al.} 2001, Astronomy and
  Astrophysics, 370, 529

\bibitem[{{Poutanen} {et~al.}(2013){Poutanen}, {Mushtukov}, {Suleimanov},
  {Tsygankov}, {Nagirner}, {Doroshenko}, \& {Lutovinov}}]{poutanen}
{Poutanen}, J., {Mushtukov}, A.~A., {Suleimanov}, V.~F., {et~al.} 2013, \apj,
  777, 115

\bibitem[{{Robba} {et~al.}(2001){Robba}, {Burderi}, {Di Salvo}, {Iaria}, \&
  {Cusumano}}]{Robba2001}
{Robba}, N.~R., {Burderi}, L., {Di Salvo}, T., {Iaria}, R., \& {Cusumano}, G.
  2001, \apj, 562, 950

\bibitem[{{Sasaki} {et~al.}(2012){Sasaki}, {M{\"u}ller}, {Kraus}, {Ferrigno},
  \& {Santangelo}}]{Sasaki12}
{Sasaki}, M., {M{\"u}ller}, D., {Kraus}, U., {Ferrigno}, C., \& {Santangelo},
  A. 2012, \aap, 540, A35

\bibitem[{{Sch{\"o}nherr} {et~al.}(2014){Sch{\"o}nherr}, {Schwarm}, {Falkner},
  {Dauser}, {Ferrigno}, {K{\"u}hnel}, {Klochkov}, {Kretschmar}, {Becker},
  {Wolff}, {Pottschmidt}, {Falanga}, {Kreykenbohm}, {F{\"u}rst}, {Staubert}, \&
  {Wilms}}]{Schoenherr2014}
{Sch{\"o}nherr}, G., {Schwarm}, F.-W., {Falkner}, S., {et~al.} 2014, \aap, 564,
  L8

\bibitem[{Smith \& Takeshima(1998{\natexlab{a}})}]{Smith:1998p3835}
Smith, D.~A. \& Takeshima, T. 1998{\natexlab{a}}, IAU Circ., 7014, 1

\bibitem[{Smith \& Takeshima(1998{\natexlab{b}})}]{Smith:1998p2184}
Smith, D.~A. \& Takeshima, T. 1998{\natexlab{b}}, The Astronomer's Telegram,
  36, 1

\bibitem[{{Tanaka}(1986)}]{Tanaka1986}
{Tanaka}, Y. 1986, in Lecture Notes in Physics, Berlin Springer Verlag, Vol.
  255, IAU Colloq. 89: Radiation Hydrodynamics in Stars and Compact Objects,
  ed. D.~{Mihalas} \& K.-H.~A. {Winkler}, 198

\bibitem[{Titarchuk(1994)}]{Titarchuk:1994p6894}
Titarchuk, L. 1994, Astrophysical Journal, 434, 570

\bibitem[{Verrecchia {et~al.}(2002)Verrecchia, Israel, Negueruela, Covino,
  Polcaro, Clark, Steele, Gualandi, Speziali, \& Stella}]{Verrecchia:2002p2366}
Verrecchia, F., Israel, G.~L., Negueruela, I., {et~al.} 2002, Astronomy and
  Astrophysics, 393, 983

\bibitem[{{White} {et~al.}(1983){White}, {Swank}, \& {Holt}}]{White1983}
{White}, N.~E., {Swank}, J.~H., \& {Holt}, S.~S. 1983, \apj, 270, 711

\bibitem[{Wilson {et~al.}(2003)Wilson, Finger, Coe, \&
  Negueruela}]{Wilson:2003p2397}
Wilson, C.~A., Finger, M.~H., Coe, M.~J., \& Negueruela, I. 2003, The
  Astrophysical Journal, 584, 996

\bibitem[{Wilson {et~al.}(1998)Wilson, Finger, Wilson, \&
  Scott}]{Wilson:1998p3746}
Wilson, C.~A., Finger, M.~H., Wilson, R.~B., \& Scott, D.~M. 1998, IAU Circ.,
  7014, 2

\bibitem[{Woo {et~al.}(1996)Woo, Clark, Levine, Corbet, \&
  Nagase}]{Woo:1996p1109}
Woo, J.~W., Clark, G.~W., Levine, A.~M., Corbet, R. H.~D., \& Nagase, F. 1996,
  Astrophysical Journal v.467, 467, 811

\end{thebibliography}
\end{document}